\documentclass[showpacs,preprintnumbers,amsmath,amssymb]{revtex4}

\usepackage{graphicx}
\usepackage{dcolumn}
\usepackage{bm}
\begin{document}
\title{Conditional implementation of a quantum phase gate with distant atoms trapped in different optical cavities}
\author{XuBo Zou, and W. Mathis }
\address{Electromagnetic Theory Group at THT,\\
Department of Electrical Engineering, University of Hannover,
Germany}
\begin{abstract}
We propose a scheme for conditional implementation of a quantum
phase gate by using distant atoms trapped in different optical
cavities. Instead of direct interaction between atoms, the present
scheme makes use of quantum interference of polarized photons
decaying from the optical cavities to conditionally create the
desired quantum phase gate between two distant atoms. The proposed
scheme only needs linear optical elements and a two-fold
coincidence detection, and are insensitive to the quantum noise.
The scheme can be directly used to prepare any quantum state of
many distant atoms.

\end{abstract}
\pacs{03.65.Ud, 42.50.-p}
 \maketitle
The manipulation and control of quantum entanglement between many
distant atoms is an important challenge for implementation of
quantum communication and computation\cite{quan}. Recently,
several physical systems have been suggested as possible
candidates for engineering quantum entanglement: cavity QED
systems \cite{tc}, trapped ion systems \cite{cd},quantum dot
systems \cite{loss} and Josephson-junction device
systems\cite{mak}. Among them, cavity QED systems are paid more
attention. This is due to the fact that cold and localized atoms
are well suitable for storing quantum information in long-lived
internal states, and the photons are natural source for fast and
reliable transport of quantum information over long distance. A
number of scheme have been proposed for manipulating quantum
entanglement between atoms and cavity fields\cite{qed,entang}. In
particular, entangled states of two or three particles have been
demonstrated experimentally in high-Q cavities \cite{ar1,ar2}.
Schemes of this type are based on the controlling of direct or
indirect interaction between the atoms, which are intended to be
entangled. Since most of these schemes require a high-Q cavity
field, decoherence caused by cavity decay can be neglected.

Conditional measurements offer another possible way to engineer
quantum entanglement and implement quantum information processing.
In Ref\cite{zoller}, Cabrillo et al proposed a scheme to entangle
quantum states of spatial widely separated atoms by weak laser
pulses and the detection of the subsequent spontaneous emission.
Further, several schemes have also been proposed for generating
the EPR state between two distant atoms trapped in different
optical cavities\cite{bose,bose1}. In Ref\cite{zou}, schemes were
also proposed for generating quantum entanglement of many distant
atoms. However these schemes are not easily extended to generate
any quantum entanglement. In order to generate any states and
implement quantum algorithm on demand, it is necessary to propose
a scheme to implement quantum logic gates. The universality of the
set of the controlled-NOT gate ( or controlled-phase gate ) and
single-qubit gates implies that we can create any states and
implement any quantum algorithm by constructing a quantum circuit
using these gates. In Ref\cite{pro}, Protsenko et al proposed a
scheme for conditional implementation of a quantum controlled-NOT
gate between two distant atoms by the detection of single
scattered photon. A disadvantage of this schemes is to require
single-photon detector to distinguish zero photon, one photon and
more than one photon. The inefficiency of photon detector
influences the fidelity of the quantum operations. Another
disadvantage of the scheme is to require that the laser pulses
have to be sufficiently weak to ensure that the probability of
exciting both atoms is much smaller the probability to excite only
one atom. It such case, it will be a time-consuming task to detect
single scattered photon. In this paper, we propose an alternative
scheme to implement conditional quantum phase gate with two
distant atoms trapped in different optical cavities with the
two-fold coincidence detection. The protocol has the following
favorable features. (1) The scheme is insensitive to the
imperfection of the photon detectors, i.e. the scheme does not
require distinguishing between zero, one and two photons. (2) The
scheme is insensitive to the phase accumulated by the photons on
their way from the cavities to the place where they are detected.
(3)The scheme does not the weakly driven laser pulses as both
atoms are excited simultaneously. (4) The scheme can be used to
prepare any quantum entanglement of many distant atoms. The
quantum entanglement between many distant sites is an essential
element of quantum network construction.

The schematic representation of our scheme for conditional quantum
phase gate is shown in Fig.1, which consists of two identical
atoms 1 and 2 confined separately in two optical cavities 1 and 2,
respectively. The photons leaking out from the cavities 1 and 2
are interfered at a polarization beam splitter (PBS), with the
outputs detected by four single-photon detectors after two quarter
wave plates ( QWP ) and two polarization beam splitters. For the
photons decaying from the cavity 1, a quarter wave plate is
inserted before the first polarization beam splitter. As shown in
Fig.2, each atom (index $j=1,2$) has four Zeeman sublevels
$|g_H\rangle_j$, $|g_V\rangle_j$, $|s_H\rangle_j$, $|s_V\rangle_j$
and two excited state $|e_H\rangle_j$, $|e_V\rangle_j$. The
lifetime of the atomic levels $|g_H\rangle_j$, $|g_V\rangle_j$,
$|s_H\rangle_j$, $|s_V\rangle_j$ is assumed to be comparatively
long so that spontaneous decay of these states can be neglected.
We encode the ground states $|g_H\rangle_j$ and $|g_V\rangle_j$ as
logic zero and one states, i.e. $|g_H\rangle_j=|0\rangle_{j}$ and
$|g_V\rangle_j=|1\rangle_{j}$. Such an atomic level structure has
been proposed to implement quantum computation in a single
cavity\cite{cirac}. The transitions
$|g_{H}\rangle_j\Longleftrightarrow|e_H\rangle_j$ and
$|g_V\rangle_j\Longleftrightarrow|e_V\rangle_j$ are coupled to two
degenerate cavity modes $a_{jH}$ and $a_{jV}$ with different
polarization $H$ and $V$, respectively. The transitions
$|e_H\rangle_j\Longleftrightarrow|s_H\rangle_j$ and
$|e_V\rangle_j\Longleftrightarrow|s_V\rangle_j$ are driven by two
classical fields, respectively. We assume that the classical laser
fields and the cavity fields are detuned from their respective
transitions by the same amount. In the case of large detuning the
excited states can be eliminated adiabatically to obtain the
effective interaction Hamiltonian (in the interaction picture)
$$
H_j=\Omega(a_{jH}|g_{H}\rangle_j{_j}\langle s_{H}|
+a_{jH}^{\dagger}|s_{H}\rangle_j{_j}\langle g_{H}|
+a_{jV}|g_{V}\rangle_j{_j}\langle s_{V}|
+a_{jV}^{\dagger}|s_{V}\rangle_j{_j}\langle g_{V}|) \eqno{(1)}
$$
where $a_{jH}$ and $a_{jV}$ ( $a_{jH}^{\dagger}$ and
$a_{jV}^{\dagger}$ ) are the annihilation ( creation ) operators
of the $H$ and $V$ polarization modes of the $j$th cavity. Here we
assumed that the effective coupling constants of the atoms coupled
with their cavity modes are same, which are described by $\Omega$.
In order to investigate the quantum dynamics of the system, it is
convenient to follow a quantum trajectory description \cite{jump}.
The evolution of the system's wave function is governed by a
non-Hermitian Hamiltonian
$$
H_j^{\prime}=H_j-i\kappa(a_{jH}^{\dagger}a_{jH}+a_{jV}^{\dagger}a_{jV})
\eqno{(2)}
$$
as long as no photon decays from the cavity. Here we assume that
two optical cavities have the same loss rate $\kappa$ for the all
modes. If a single-photon detector $D_j$ ($ j=1,2,3,4$) detects a
photon, the coherent evolution according to $H_j^{\prime}$ is
interrupted by a quantum jump. This corresponds to a quantum jump,
which can be formulated with the operators $b_j$ on the joint
state vectors of two atom-cavity systems
$$
b_1=\frac{1}{2}(a_{1H}+a_{1V})+\frac{1}{\sqrt{2}}a_{2V}
,~~b_2=\frac{1}{2}(a_{1H}+a_{1V})-\frac{1}{\sqrt{2}}a_{2V}
$$
$$
b_3=\frac{1}{2}(a_{1H}-a_{1V})+\frac{1}{\sqrt{2}}a_{2H}
,~~b_4=-\frac{1}{2}(a_{1H}-a_{1V})+\frac{1}{\sqrt{2}}a_{2H}
\eqno{(3)}$$ In the following, we analyze the scheme in detail. In
order to demonstrate the conditional implementation of quantum
phase gate, we assume that atom, trapped in the $j$-th ( $j=1,2$ )
optical cavity, are initially prepared in the state
$$
\alpha_j|g_{H}\rangle_j+\beta_j|g_{V}\rangle_j,\eqno{(4)}
$$
and both polarization modes of optical cavity are prepared in the
vacuum states $|0,0\rangle_j$, where $|m,n\rangle_j$ denotes m
photons in polarization mode and n photons in the polarization
mode. Now we switch on the Hamiltonian(1) in each atom-cavity
system for a time $\tau$. If no photon is emitted from the cavity,
the $j$th atom-cavity system is governed by the interaction
$H_j^{\prime}$. In this case the atom-cavity state evolves to the
entangled state
$$
|\Psi\rangle_j=\alpha_j[a|g_{H}\rangle_j|0,0\rangle_{j}-ib|s_H\rangle_j|1,0\rangle_j]
+\beta_j[a|g_{V}\rangle_j|0,0\rangle_{j}-ib|s_V\rangle_j|0,1\rangle_j]
\eqno{(5)} $$ with
$$
a=\frac{\cos(\Omega_{\kappa}\tau)-\frac{\sin(\Omega_{\kappa}\tau)\kappa}{2\Omega_{\kappa}}}
{\sqrt{[\cos(\Omega_{\kappa}\tau)-\frac{\sin(\Omega_{\kappa}\tau)\kappa}{2\Omega_{\kappa}}]^2
+\frac{\sin^2(\Omega_{\kappa}\tau)\Omega^2}{\Omega_{\kappa}^2}}}
$$
$$
b=\frac{\sin(\Omega_{\kappa}\tau)\Omega}{\Omega_{\kappa}\sqrt{[\cos(\Omega_{\kappa}\tau)
-\frac{\sin(\Omega_{\kappa}\tau)\kappa}{2\Omega_{\kappa}}]^2
+\frac{\sin^2(\Omega_{\kappa}\tau)\Omega^2}{\Omega_{\kappa}^2}}}
$$
$$
\Omega_{\kappa}=\sqrt{\Omega^2-\kappa^2/4} \eqno{(6)}
$$
The probability that no photon is emitted during this evolution
becomes
$$P_{single}=e^{-\kappa\tau}\{[\cos(\Omega_{\kappa}\tau)-\frac{\sin(\Omega_{\kappa}\tau)\kappa}{2\Omega_{\kappa}}]^2
+\frac{\sin^2(\Omega_{\kappa}\tau)\Omega^2}{\Omega_{\kappa}^2}\}\eqno{(7)}
$$
We assume that the interaction Hamiltonian (1) is applied to each
atom-cavity system simultaneously, so that the atom-cavity states
$|\Psi\rangle_j$ ends at the same time. This implements the first
step of the protocol. The probability that this stage is a success
is the probability that no photon decays from either atom-cavity
system during the preparation. This quantity is given by
$P_{suc}=P_{single}^2$.

Next we consider the second step of the scheme, in which we make a
photon number measurement with four single-photon detectors $D_j$
($j=1,2,3,4$) on the output modes of the setup. In this step, the
joint state of two atom-cavity systems becomes prepared in the
form
$$
|\Phi(0)\rangle=|\Psi\rangle_1\otimes|\Psi\rangle_2\eqno{(8)} $$
where the state $|\Psi\rangle_j$ is given by Eq.(5). We assume
that photons are detected at the time $t$. This assumption is
posed to calculate the system's time evolution during this time
interval in a consistent way with the
"no-photon-emission-Hamiltonian" (2). The joint state of the total
system evolves into
$$
|\Phi(t)\rangle=|\Psi(t)\rangle_1\otimes|\Psi(t)\rangle_2
\eqno{(9)} $$ with
$$
|\Psi(t)\rangle_j= \frac{1}{a^2+b^2e^{-2\kappa{t}}}
\{\alpha_j[a|g_{H}\rangle_j|0,0\rangle_{j}-ibe^{-\kappa{t}-i\varphi_j}|s_H\rangle_j|1,0\rangle_j]
+\beta_j[a|g_{V}\rangle_j|0,0\rangle_{j}-ibe^{-\kappa{t}-i\varphi_j}|s_V\rangle_j|0,1\rangle_j]\}
\eqno{(10)} $$ where the phases $\varphi_j=kL_j$, $k$ is the wave
number and $L_{j}$ are the optical lengths which photons travel
through the optical system towards the photon detectors. The
detection of one photon with the detector $D_j$ corresponds to a
quantum jump, which can be formulated with the operator $b_j$ on
the joint state $|\Phi(t)\rangle$. If $D_1$ and $D_3$ detect one
photon at nearly the same time and $D_2$ and $D_4$ do not detect
any photon, or vice, the state of the total system is projected
into
$$
\alpha_1\alpha_2|s_{H}\rangle_1|s_{H}\rangle_2+\beta_1\alpha_2|s_{V}\rangle_1|s_{H}\rangle_2
+\alpha_1\beta_2|s_{H}\rangle_1|s_{V}\rangle_2-\beta_1\beta_2|s_{V}\rangle_1|s_{V}\rangle_2
\eqno{(11)}
$$
If $D_1$ and $D_4$ detect one photon at nearly the same time and
$D_2$ and $D_3$ do not detect any photon during that time
interval, or vice, the state of the total system becomes projected
into
$$
\alpha_1\alpha_2|s_{H}\rangle_1|s_{H}\rangle_2+\beta_1\alpha_2|s_{V}\rangle_1|s_{H}\rangle_2
-\alpha_1\beta_2|s_{H}\rangle_1|s_{V}\rangle_2+\beta_1\beta_2|s_{V}\rangle_1|s_{V}\rangle_2
\eqno{(12)}
$$
where we have neglected the multiplicative factor
$e^{i(\varphi_1+\varphi_2)}$ in Eq.(11) and (12). This result
demonstrates that phase accumulated by the photons have no effect
on the conditional implementation of the quantum operation. Based
on the result of measurement, we can apply fast Raman transitions
to individually manipulate the atoms 1 and 2, and map the states
Eq.(11) or Eq.(12) to the state
$$
\alpha_1\alpha_2|g_{H}\rangle_1|g_{H}\rangle_2+\beta_1\alpha_2|g_{V}\rangle_1|g_{H}\rangle_2
+\alpha_1\beta_2|g_{H}\rangle_1|g_{V}\rangle_2-\beta_1\beta_2|g_{V}\rangle_1|g_{V}\rangle_2
\eqno{(13)}
$$
which demonstrates the conditional implementation of quantum
controlled phase gate
$$
|g_{H}\rangle|g_{H}\rangle\rightarrow|g_{H}\rangle|g_{H}\rangle,
~~~|g_{H}\rangle|g_{V}\rangle\rightarrow|g_{H}\rangle|g_{V}\rangle
$$
$$
|g_{V}\rangle|g_{H}\rangle\rightarrow|g_{V}\rangle|g_{H}\rangle,
~~~|g_{V}\rangle|g_{V}\rangle\rightarrow-|g_{V}\rangle|g_{V}\rangle
\eqno{(14)}
$$
The success probability of the scheme is
$P_s=b^4(1-e^{-2\kappa{t}})^2/2$. We now give a brief discussion
on the influence of practical noise on the scheme. In the present
scheme, the lifetime of the atomic ground levels is assumed to be
comparatively long so that spontaneous decay of these states can
be neglected. The dominate noise of the scheme is the photon loss,
which includes the contribution from channel attenuation, and the
inefficiency of the single-photon detectors. All these kinds of
noise can be considered by an overall photon loss probability
$\eta$\cite{bose}. It is noticed that the present scheme is based
on the two-photon coincidence detection. If one photon is lost, a
click from each of the detectors is never recorded. In this case,
the scheme fails to generate the expected quantum operation.
Therefore the photon loss only decreases the success probability
$P_s$ by a factor of $(1-\eta)^2$, but have no influence on the
fidelity of the expected operation.

In summary, we have proposed a scheme to implement the conditional
quantum phase gate for two distant atoms trapped in different
optical cavities by using interference of polarization photons.
Compared with previous scheme\cite{pro}, the scheme is insensitive
to the imperfection of the photon detectors. The detection
efficiency does not affect the fidelity of the conditional quantum
operation and photon counting detectors are not needed. Low
detection efficiency lowers only the success probability. The most
possible application of the present scheme is to create quantum
entanglement between many distant atoms trapped in different
optical cavities. Recently, researchers are turning their interest
towards characterizing and generating multipartite entanglement,
and using it for more general and useful applications\cite{milti}.
Although several schemes have been proposed for quantum
entanglement of many distant atoms\cite{zou}, these schemes are
not easily extended to generate any quantum entanglement. By
employing the quantum logic networks proposed in
Ref\cite{general}, the present scheme can be directly used to
conditionally generate any quantum state of many distant atoms.

\begin{flushleft}

{\Large \bf Figure Captions}

\vspace{\baselineskip}

{\bf Figure 1.} The schematic setup for conditional quantum phase
gate of two distant atoms trapped in different optical cavities,
which includes three polarization beam splitters (PBS), three
quarter wave plates ( QWP ) and four single-photon detectors.

{\bf Figure 2.} The relevant level structure with ground state
$|g_H\rangle$, $|s_H\rangle$ , $|g_V\rangle$, $|s_V\rangle$ and
excited states $|e_H\rangle$, $|e_V\rangle$. The transitions
$|e_H\rangle\rightarrow|s_H\rangle$ and
$|e_V\rangle\rightarrow|s_V\rangle$ are coupled by the classical
lasers and the transitions $|e_H\rangle\rightarrow|g_H\rangle$ and
$|e_V\rangle\rightarrow|g_V\rangle$ are coupled to two degenerate
cavity modes $a_{H}$ and $a_{V}$ with different polarization $H$
and $V$.
\end{flushleft}

\end{document}